\documentclass[aps,prb,12pt,a4paper,superscriptaddress,reprint]{revtex4-1}

\usepackage[utf8]{inputenc}
\usepackage{graphicx}
\usepackage{bm}
\usepackage{slashed}
\usepackage{amsmath}
\usepackage{amsfonts}
\usepackage{amssymb}
\usepackage{xcolor}
\usepackage{ulem}
\usepackage{dcolumn}
\providecommand{\U}[1]{\protect\rule{.1in}{.1in}}
\DeclareMathOperator{\Lame}{\hat{\mathcal{L}}}
\providecommand{\U}[1]{\protect\rule{.1in}{.1in}}
\providecommand{\U}[1]{\protect\rule{.1in}{.1in}}
\newcommand{\be}{\begin{equation}}
\newcommand{\en}{\end{equation}}

\begin{document}

\title{Theory of magnetoelastic resonance in a mono-axial chiral helimagnet}

\author{A.~A. \surname{Tereshchenko}}
\affiliation{Institute of Natural Sciences and Mathematics, Ural Federal University, Ekaterinburg 620002, Russia}

\author{A.~S. \surname{Ovchinnikov}}
\affiliation{Institute of Natural Sciences and Mathematics, Ural Federal University, Ekaterinburg 620002, Russia}
\affiliation{Institute of Metal Physics, Ural Division of the Russian Academy of Sciences, Ekaterinburg 620219, Russia}

\author{Igor \surname{Proskurin}}
\email{Igor.Proskurin@umanitoba.ca}
\affiliation{Institute of Natural Sciences and Mathematics, Ural Federal University, Ekaterinburg 620002, Russia}
\affiliation{Department of Physics and Astronomy, University of Manitoba, Winnipeg MB R3T 2N2, Canada}
\affiliation{Chirality Research Centre, Hiroshima University, Higashi-Hiroshima, Hiroshima 739-8526, Japan}

\author{E.~V. \surname{Sinitsyn}}
\affiliation{Institute of Natural Sciences and Mathematics, Ural Federal University, Ekaterinburg 620002, Russia}

\author{Jun-ichiro \surname{Kishine}}
\email{kishine@ouj.ac.jp}
\affiliation{Division of Natural and Environmental Sciences, The Open University of Japan, Chiba 261-8586, Japan}

\date{\today}

\begin{abstract}
We study magnetoelastic resonance phenomena in a mono-axial chiral helimagnet belonging to hexagonal crystal class. By computing the spectrum of coupled elastic wave and 
spin wave, it is demonstrated how hybridization occurs depending on their chirality. 
Specific features of the magnetoelastic resonance are
discussed for the conical phase and the soliton lattice phase stabilized in the
mono-axial chiral helimagnet. The former phase exhibits appreciable
non-reciprocity of the spectrum, the latter is characterized by a
multi-resonance behavior. 
We propose that the non-reciprocal spin wave around the forced-ferromagnetic state has potential capability to convert the linearly polarized elastic wave to circularly polarized one with the chirality opposite to the spin wave chirality.
\end{abstract}


\maketitle

\section{Introduction}

Many recent studies have focused on physical properties of chiral
helimagnets (CHM). It is {widely} recognized that coupling of {lattice degrees
of freedom} with magnetism plays a significant role in this class of materials. For example,  the cubic chiral helimagnet  MnSi\cite{Fawcett1970,Matsunaga1982} exhibits the  anomalies in the thermal expansion coefficient 
 and similarly MnGe\cite{Valkovskiy2016} exhibits magnetic peculiarities connected with distortion of the B20 structure upon heating.

Magnetoelastic interaction may contribute either to dynamic elastic
deformations that affect significantly the dynamics of magnetic moments or to
static strains, which in turn influence the dispersion and band-gaps of
the coupled magnetoelastic waves. This coupling was {argued} in relation to a
possible structural transition in Mn${}_{1-x}$Fe${}_{x}$Ge solid solutions
\cite{Makarova2012,Dyadkin2014,Martin2016}. Early theoretical studies of the
magnetoelastic interaction in cubic helimagnets with B20 {structure} predicted
an appearance of non-analytical wave-vector dependence for the static
susceptibility as a result of magnetization-induced inhomogeneous strains
\cite{Plumer1982,Plumer1984}; it was demonstrated that this interaction tends
to disrupt the assumed helical structure \cite{Maleyev2009}.

One of the powerful tools to investigate specific features of the
magnetoelastic coupling are ultrasound measurements, where characteristics of
propagation of high-frequency elastic waves are indicated by a dependence of
the velocity and attenuation of the ultrasonic waves on magnetic properties of
the solid. They are reputed to be a valuable probe to investigate magnetic
phase transitions in MnSi due to high sensitivity and accuracy
\cite{Petrova2009,Petrova2016}. Sound velocities measured in these studies are
highly sensitive to local values of elastic constants and their evaluation
does not involve any sophisticated experimental technique.

One of the most important reasons of keen interest in chiral helimagnets is
driven by the unique soliton-like forms of magnetic order revealed in these
materials: the chiral soliton lattice {(CSL)} actually observed in {CrNb$_{3}%
$S$_{6}$} \cite{Togawa2012} and the skyrmion lattice found, for example, in
MnSi, (Fe,Co)Si and Cu${}_{2}$OSeO${}_{3}$
\cite{Binz2009,Yu2010,Adams2010,Seki2012}. Ultrasonic measurements being
compared with magnetic and electric ones demonstrate clear advantages for
exploring these topological objects: they are not restricted by electric
conductivity of a material; due to magnetoelastic interaction, they provide
insight into anisotropic properties of the magnetic lattices by comparing
different elastic modes; lastly, they make possible to determine directly
elasticity and viscosity of these lattices as a result of the magnetoelastic
coupling. Mechanical control of the skyrmion lattice phase demonstrated in a
bulk MnSi single crystal is of considerable interest; it is achieved with a
mechanical stress and a low energy cost \cite{Nii2015}. Deep understanding of
the issue is vital for potential applications in technology.

A growing interest in the nontrivial topological phases of the chiral
helimagnets dictates an urgent need to elaborate an appropriate formalism of
the magnetoelastic interaction of these materials. The seminal theory of
magnetoelastic waves in ferromagnetic crystals, originally suggested by Kittel
\cite{Kittel1958}, has been expanded into the class of helimagnets with the
Dzyaloshinskii-Moryia (DM) exchange coupling over few decades ago
\cite{Stefanovski1969,Vlasov1973}. {However, spontaneous deformations in a
ground state were ignored in these treatments.} The theory developed in Ref.
\cite{Turov1983} {overcame} this drawback; a pertinent investigation for the
conical phase of the relativistic spiral has been later reported
\cite{Shavrov1989,Bychkov1990}. Recently this problem has been under new
scrutiny in the light of  of magnetoelectric hexaferrites, where the magnetoelastic
resonance is largely the same as for the phase of forced ferromagnetism in the
monoaxial CHM \cite{Vittoria2015}.  We also point out a remarkable feature of spin wave propagation in the conical phase in chiral helimangets. A preferable spin-wave helicity (left-handed or right-handed) is fixed by the DM interaction. Consequently, non-reciprocal magnon transport is realized.\cite{Iguchi2015,Seki2015}

 {The coupling between
acoustic phonons and magnons was incorporated} to explore {the effects of the
spin-lattice coupling in} the topologically nontrivial skyrmion lattice in
MnSi and MnGe \cite{Zhang2017}. The magnetoelastic interaction results from
expanding the strengths of both Heisenberg exchange interaction and the
Dzyaloshinskii-Moriya interaction up to the linear order of phonon degrees of
freedom. Efficiency of such a form of the magnetoelastic coupling was
experimentally demonstrated for the skyrmion lattice in MnGe, where the
elastic response is an order of magnitude larger than the conventional case
(for example, in MnSi) was reported \cite{Kanazawa2016}. To calculate
ultrasonic {responses} in MnSi the thermodynamical model was {used}
\cite{Hu2017}, which incorporates a magnetoelastic functional with necessary
high-order interactions allowed by group theory. Unfortunately, a progress in
this direction is severely hampered by lack of a generally accepted
theoretical model for the skyrmion lattice phase \cite{Rosch2016}.

In this paper, we fill a gap coming from, to the best of our knowledge, an
absence of a theory of magnetoelastic interactions in the chiral soliton
lattice. This case is certainly of a special interest: a control of the
period of the soliton lattice by means of an external magnetic field enables
governing a resonant frequency in a substantial way. Our analysis is intended
for crystals of the hexagonal symmetry which the real prototype compound
{CrNb$_{3}$S$_{6}$} belongs to. Until now, only the case of the exchange
spiral has been investigated for this symmetry \cite{Bychkov1990}. A {spiral}
magnetic order owing to the DM interaction was previously analyzed for a media
with isotropic elastic and magnetoelastic properties \cite{Shavrov1993} that
can be applied to the chiral magnetic materials of cubic symmetry, MnSi and
FeGe. The aim of our investigation is to find out specific features of
magnetoelastic resonance in the magnetic soliton lattice and to provide
insight into factors that affect the process significantly. In addition, we
revisit a case of the conical phase to discuss salient non-reciprocity effects
in propagation of magnetoelastic waves.

The paper is organized as follows. In Sec. II, the model of the interaction
between the magnetic and the elastic degrees of freedom is formulated. Sec.
III provides a treatment of the magnetostriction problem, i.e., a calculation
of elastic deformations caused by magnetization of the soliton lattice. In
Sec. IV, the coupled system of dynamical equations for the lattice and the
spin variables is solved; the spectrum of the magnetoelastic waves is
analyzed. For the sake of simplicity, we consider the waves traveling along a
principal axis of the crystal. In Sec. V, the conclusions are presented.

\section{The model}

\begin{figure}[ptb]
\includegraphics[scale=0.35]{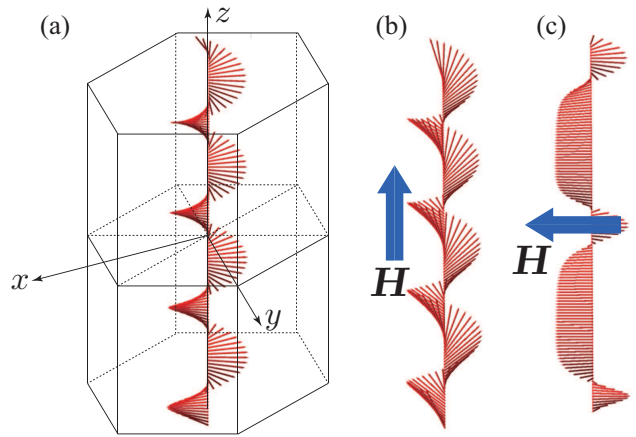} \caption{(color online)
The coordinate system for a hexagonal crystal with a magnetic spiral inside
used throughout the paper (a); schematic pictures of the magnetic ordering for
the conical phase (b) and the soliton lattice phase (c).}%
\label{Fig1}%
\end{figure}

We consider a hexagonal chiral helimagnet, where a modulated magnetic ordering
characterized by the magnetization $\boldsymbol{M}(z,t)$, is stabilized along
the symmetry direction taken further as $z$-axis. In hexagonal crystals, the
total energy density, which takes into account interaction with elastic
deformations, can be expressed in the following form
\begin{equation}
\mathcal{F}=\frac{J}{2}\left(  \partial_{z}\boldsymbol{M}\right)  ^{2}%
+D\hat{z}\cdot\left[  \boldsymbol{M}\times\partial_{z}\boldsymbol{M}\right]
-\boldsymbol{H}\cdot\boldsymbol{M}+\mathcal{F}_{ME}+\mathcal{F}_{E}%
,\label{FreeEn}%
\end{equation}
where the first term is that of the Heisenberg model with the ferromagnetic
exchange coupling $J$; the second terms is the DM
interaction of the strength $D$, and the third one describes the interaction
of the magnetization with the external magnetic field $\boldsymbol{H}$. The last
two terms stand for the magnetoelastic and elastic energy densities
respectively, whose explicit form for the hexagonal crystal structure is given
by \cite{Mason1954}
\begin{align}
\mathcal{F}_{E}&=\frac{c_{11}}{2}\left(  u_{xx}^{2}+u_{yy}^{2}\right)
+\frac{c_{33}}{2}u_{zz}^{2}+\left(  c_{11}-c_{12}\right)  u_{xy}%
^{2}\nonumber\\
&+c_{12}u_{xx}u_{yy}+2c_{44}\left(  u_{xz}^{2}+u_{yz}^{2}\right)
+c_{13}\left(  u_{xx}+u_{yy}\right)  u_{zz},\label{FE}%
\end{align}%
\begin{align}
\mathcal{F}_{ME}&=\left(  b_{11}-b_{12}\right)  \left(  u_{xx}M_{x}^{2}%
+2u_{xy}M_{x}M_{y}+u_{yy}M_{y}^{2}\right)  \nonumber\\
&+\left(  b_{13}-b_{12}\right)  \left(  u_{xx}+u_{yy}\right)  M_{z}^{2}+\left(
b_{33}-b_{31}\right)  u_{zz}M_{z}^{2}\nonumber\\
&+2b_{44}\left(  u_{xz}M_{x}M_{z}+u_{yz}M_{y}M_{z}\right)  ,\label{FME}%
\end{align}
where $u_{ij}$ is the deformation tensor defined in terms of elastic
deformations $s_{i}$
\begin{equation}
u_{ij}=\frac{1}{2}\left(  \frac{\partial s_{i}}{\partial x_{j}}+\frac{\partial
s_{j}}{\partial x_{i}}\right)  .
\end{equation}
where $i,j=x,y,z$ indicate directions schematically shown in Fig.~\ref{Fig1}%
~(a), and $b_{ij}$ and $c_{ij}$ are correspondingly the magnetoelastic  and the
elastic stiffness modulus constants.

To study the magnetoelastic resonance, we consider the coupled equations of
motion for $\boldsymbol{M}$ and $u_{ij}$
\begin{align}
\rho\frac{\partial^{2}s_{i}}{\partial t^{2}}  &  =\frac{\partial\sigma_{ij}%
}{\partial x_{j}},\label{EM1}\\
\frac{\partial\boldsymbol{M}}{\partial t}  &  =-\gamma\boldsymbol{M}\times
\boldsymbol{H}_{\mathrm{eff}}, \label{EM2}%
\end{align}
where both the effective field $\boldsymbol{H}_{\mathrm{eff}}=-\delta
\mathcal{F}/\delta\boldsymbol{M}$ and the stess tensor $\sigma_{ij}=\left(
1+\delta_{ij}\right)  /2\left(  \partial\mathcal{F}/\partial u_{ij}\right)
$\cite{Comstock1963} are defined by the energy density in Eq.~\ref{FreeEn},
$\rho$ is the crystal {mass}\ density, and $\gamma$ denotes the gyromagnetic
ratio. For numerical estimations later on, we use the crystallographic data
for {CrNb$_{3}$S$_{6}$} compound, which contains 20 atoms per unit cell:
twelve S atoms, six Nb atoms and two intercalated Cr atoms. The unit cell
parameters are $a=5.741\mathring{A}$ and $c=12.101\mathring{A}$ that yields
$\rho=5.029$ g/cm${}^{3}$. \cite{Mandrus2013} For a numerical value of the
magnetization $M_{0}$, we use the result $3.2\,\mu_{B}/$Cr and the nearest
Cr-Cr distance in the $ab$-plane (5.741 $\mathring{A}$) and along the $c$-axis
(6.847 $\mathring{A}$). \cite{Mandrus2013} This gives $M_{0}=131.5$ kA/m =
1649 Gs.

At present, it is hard to give any precise numerical values for the
coefficients $b_{ij}$ and $c_{ij}$ in the prototype compound, chiral
helimagnet {CrNb$_{3}$S$_{6}$}. Instead, the values of the stiffness moduli
for the parent matrix NbS${}_{2}$ of the same hexagonal structure are used:
$c_{11}=148$ GPa, $c_{12}=51$ GPa, $c_{13}=1$ GPa, $c_{33}=c_{44}=2$ GPa
\cite{Gaillac2016}. The constants $b_{ij}M_0^2$ of the order $1-10\,\text{MPa}$ are
used for estimations whenever it is necessary.

We emphasize, that in order to develop a linear theory of magnetoelastic
resonance in systems with inhomogeneous magnetization profile, it is important
to take into account the magnetostricrive effect from the magnetization
background \cite{Turov1983}, which results into the inhomogenious deformation
fied $u^{(0)}_{ij}(\boldsymbol{r})$ in the ground state induced by the
spontaneous magnetization $\boldsymbol{M}_{0}(\boldsymbol{r})$. Interestingly, as
it was pointed out in Ref.~[\onlinecite{Turov1983}], this effect of
spontaneous symmetry breaking caused by magnetic ordering in a system of the
two coupled fields is analogous to the Higgs effect in the theory of
elementary particles \cite{Higgs1964}. The spatial dependence of the
background magnetization also requires modification of methods used in
previous studies of ferromagnetic materials. For example, nonunifrom strains
can make all the magnetoelastic waves to be massless Goldstone's modes, i.e.,
in contrast to ferromagnets, no magnetoelastic gap appears. \cite{Shavrov1989}

\section{Magnetoelastic effect}

Previous studies of magnetoelastic waves in crystals with helicoidal magnetic order, motivated mostly by available at that time experimental data on
ultrasound excitations in rare-earth metals \cite{Shavrov1994} where the
spiral ordering originates from the competition between the exchange
couplings, demonstrated that the modulated magnetization of the ground state
results in nonuniform equilibrium deformations of the crystal
\cite{Shavrov1989}. The results for the cubic crystals with a relativistic
spiral structure stabilized by the DM interaction were
addressed in Ref.~\onlinecite{Shavrov1993}. Below, we summarize the results
for the hexagonal chiral crystals which demonstrate substantial difference from the cubic case.

At first, we briefly review different modulated magnetic phases realized in
chiral helimagnets of hexagonal symmetry under the external static magnetic
field. For this purpose, we use classical representation of the magnetization
$\boldsymbol{M}_{0}=M_{0}\left(  \sin\theta_{0}\cos\varphi_{0},\sin\theta_{0}%
\sin\varphi_{0},\cos\theta_{0}\right)  $ parametrized by the azimuthal
($\varphi$) and polar ($\theta$) angles. When the magnetic field in
Eq.~\ref{FreeEn} is applied along the $\hat{z}$-direction, the conical phase
characterized by $0<\theta_{0}<\pi/2$ and $\varphi_{0}=qz$ is stabilized for
$H^{z}<H_{c}^{z}$, as schematically shown in Fig.~\ref{Fig1}~(b), where
$q=-D/J$ is the helical pitch, $H_{c}^{z}=M_{0}D^{2}/J$ is the critical field
for the conical phase, and $\cos\theta_{0}=H^{z}/H_{c}^{z}$. For $H^{z}%
>H_{c}^{z}$ the forced ferromagnetic state along $\hat{z}$-axis appears. The
situation is completely different when $\boldsymbol{H}$ is applied
perpendicular to the chiral axis, see Fig.~\ref{Fig1}~(c). In this case, the
periodic nonlinear structure called the magnetic soliton lattice corresponds
to the minimum of magnetic energy for any nonzero $H^{x}$ and determined by
the solution of the sine-Gordon equation with $\theta_0=\pi/2$ and
\begin{equation}
\varphi_{0}(z)=\pi+2\text{am}\left(  \frac{mz}{\kappa}\right)  , \label{SLsol}%
\end{equation}
where $\text{am}(\ldots)$ is the Jacobi's amplitude function with the elliptic
modulus $\kappa$, $0\leq\kappa^{2}<1$. The parameter $m^{2}=H^{x}/JM_{0}$
plays a role of the first breather mass in the context of the sine-Gordon
model and determines the period of the soliton lattice. The modulus $\kappa$
is determined by the relation $\left(  \kappa/E\right)  ^{2}=H^{x}/H_{c}^x$,
where $H_{c}^{x}=JM_{0}\left(  \pi q/4\right)  ^{2}$ is the critical field for
the soliton lattice phase at which the incommensurate-commensurate phase
transition occurs; $E$ is the elliptic integral of the second kind. At zero
magnetic field, both the soliton lattice and the conical phases degenerate
into the simple spiral with $\varphi_0=qz$.

Having determined the magnetic background, we are in a position to study
magnetostriction effects. At this point, the approximate character of our
treatment should be highlighted. We imply that the magnetic ordering is
determined independently from the elastic subsystem by minimizing only the
magnetic part of the total energy density in Eq.~\eqref{FreeEn}. This
approach, which is justified when magnetoelastic interaction is much weaker
that magnetic interactions, allows us to determine inhomogeneous deformations
induced by magnetic background, but ignores the backward effect of elastic
subsystem on magnetic ordering. The accurate treatment should minimize the
total energy simultaneously with respect to the magnetization and elastic
deformations, which eventually leads the double sine-Gordon model, also known
as the sine-Gordon model with crystalline anisotropy of the second order
\cite{Izyumov1984}.

In order to find the induced deformation field $u_{ij}^{(0)}$, we apply the
Saint-Venant's compatibility condition for the infinitesimal strain components, which ensures that the strain is the symmetric derivative of some vector field,
\cite{Chandra1994}
\begin{equation}
\partial_{ij}^{2}u_{kl}+\partial_{kl}^{2}u_{ij}-\partial_{ik}^{2}%
u_{jl}-\partial_{jl}^{2}u_{ik}=0,
\end{equation}
where $ijkl=1212,1313,2323,1213,2123,3132$. In the present case of
one-dimensional modulation $u_{ij}=u_{ij}(z)$, it reduces to
\begin{equation}
\partial_{z}^{2}u_{xx}=\partial_{z}^{2}u_{yy}=\partial_{z}^{2}u_{xy}=0,
\end{equation}
which yields constant $u_{xx}=u_{xx}^{(0)}$, $u_{yy}=u_{yy}^{(0)}$, and
$u_{xy}=u_{xy}^{(0)}$ under the the requirement of finiteness of the
deformations. Inserting these displacements into Eqs.~(\ref{FE},\ref{FME}) and
minimizing the total energy with respect to $u_{zz}$, $u_{xz}$, and $u_{yz}$,
we find the remaining components of the deformation tensor
\begin{align}
&  u_{zz}^{(0)}=-\frac{\left(  b_{33}-b_{31}\right)  }{c_{33}}M_{0z}^{2}%
-\frac{c_{13}}{c_{33}}\left(  u_{xx}^{(0)}+u_{yy}^{(0)}\right)  ,\label{uzz}\\
&  u_{xz}^{(0)}(z)=-\frac{b_{44}}{2c_{44}}M_{0z}M_{0x}(z),\label{uxz}\\
&  u_{yz}^{(0)}(z)=-\frac{b_{44}}{2c_{44}}M_{0z}M_{0y}(z). \label{uyz}%
\end{align}
Substituting Eqs.(\ref{uzz})--(\ref{uyz}) back into Eqs.~(\ref{FE},\ref{FME}),
we obtain the energy density that depends only on $u_{xx}$, $u_{yy}$, and
$u_{xy}$. These values are obtained by minimization of $\mathcal{F}$ per
period $L$, $L^{-1}\int_{0}^{L}\mathcal{F}dz$, which eventually leads to the
results $u_{xy}^{(0)}=0$ and
\begin{align}
u_{xx}^{(0)}=u_{yy}^{(0)}=\frac{M_{0}^{2}}{\Delta}\left[  c_{13}\left(
b_{33}-b_{31}\right)  \cos^{2}\theta_{0}\right. \nonumber\\
\left.  -c_{33}\left(  b_{13}-b_{12}\right)  \cos^{2}\theta_{0}-\frac{c_{33}%
}{2}\left(  b_{11}-b_{12}\right)  \sin^{2}\theta_{0}\right]  , \label{uxx}%
\end{align}
where $\Delta=c_{33}\left(  c_{11}+c_{12}\right)  -2c_{13}^{2}$.

The equations above demonstrate that in hexagonal crystals the helical
magnetic ordering triggers the screw deformations $u_{xz}^{(0)}$,
$u_{yz}^{(0)}$ whereas the shear and the normal strains remain uniform, in
agreement with previous results for hexagonal crystals with the exchange
spiral ordering \cite{Shavrov1989}. {The presence of the screw deformations is
a remarkable feature of the mono-axial crystal classes, whereas it is absent in
the cubic classes.}\cite{Shavrov1993} Such type of hybridization between the
spin modulations and the elastic deformations supports an idea that spin
chirality is connected to the torsion deformations. This correspondence has
been proved experimentally in Ho metal, where the left-screw domain population
excess was reached after exertion of the torsion elastic deformation
\cite{Fedorov1997}. However, similar experiments were found unsuccesfull in
cubic chiral magnets, such as Fe${}_{1-x}$Co${}_{x}$Si and Mn${}_{1-x}$%
Fe${}_{x}$Si. \cite{Grigoriev2009,Grigoriev2010}

\section{magnetoelastic resonance}

The theory of linear magnetoelastic resonance follows from Eqs.~\ref{EM1}
and \ref{EM2} by expanding them near the equilibrium magnetization
$\boldsymbol{M}=\boldsymbol{M}_{0}(z)+\delta\boldsymbol{m}(z,t)$ and
deformation fields $u_{ij}=u_{ij}^{(0)}(z)+\delta u_{ij}(z,t)$ and keeping
only linear contributions in terms of small perturbations $\delta
\boldsymbol{m}(z,t)$ and $\delta u_{ij}(z,t)$. For the elastic deformations,
the explicit expression are  as follows, $u_{xx}=u_{yy}=u_{xx}^{(0)}$,
$u_{xy}=u_{xy}^{(0)}=0$, and
\begin{equation}
u_{iz}=u_{iz}^{(0)}+\frac{1}{2}
\left(\frac{\partial s_{i}}{\partial x_3}+\frac{\partial s_{3}}{\partial x_i}\right),\quad
i=1,2,3, \label{uiz}%
\end{equation}
where $(x_1,x_2,x_3)=(x,y,z)$.

Below, we consider magnetoelastic waves in two modulated magnetic phases of
the chiral helimagnet: the conical one that appears when the static magnetic
field is applied along the chiral axis, $\boldsymbol{H}=H^z\boldsymbol{e}_{z}$,
and the soliton lattice phase arising when the field is perpendicular to the
axis, $\boldsymbol{H}=H^x\boldsymbol{e}_{x}$.

\subsection{Conical phase}

The conical phase is specified by the finite cone angle angle $0<\theta
_{0}<\pi/2$, and harmonic magnetic modulation with the helical pitch $q=-D/J$.
For the following discussion it is convenient to introduce circular amplitudes
for the magnetic and elastic waves, $M_{\pm}(z,t)=M_{x}(z,t)\pm iM_{y}(z,t)$
and $s_{\pm}(z,t)=s_{x}(z,t)\pm is_{y}(z,t)$, respectively. In these
notations, the dynamical part of the magnetization becomes $\delta m_{\pm
}(z,t)=M_{0}\cos\theta_{0}e^{\pm iqz}\delta\theta(z,t)\pm iM_{0}\sin\theta
_{0}e^{\pm iqz}\delta\varphi(z,t),$ and $\delta m_{z}(z,t)=-M_{0}\sin
\theta_{0}\delta\theta(z,t)$, which after the substitution into Eqs.~\ref{EM1}
and \ref{EM2}, together with Eq.~\ref{uiz}, gives after some algebra the
following coupled equations of motion for the elastic displacements and the
magnetization
\begin{equation}
\frac{\partial^{2}s_{z}}{\partial t^{2}}=v_{l}^{2}\frac{\partial^{2}s_{z}%
}{\partial z^{2}}-\beta_{3}\sin2\theta_{0}\frac{\partial\delta\theta}{\partial
z}, \label{dszdt}%
\end{equation}
\begin{widetext}
\begin{align}
\label{dspmdt}
\frac{\partial^2 s_{\pm}}{\partial t^2} &=  v^2_t \frac{\partial^2 s_{\pm}}{\partial z^2} + \beta_1 \cos 2 \theta_0 \frac{\partial}{\partial z} \left(   e^{\pm iqz} \delta \theta \right) \pm \frac{i}{2} \beta_1 \sin 2 \theta_0 \frac{\partial}{\partial z} \left(   e^{\pm iqz} \delta \varphi \right),  \\   \label{dthetadt}
\frac{\partial \delta \theta}{\partial t} &= JM_0 \gamma  \sin \theta_0 \frac{\partial^2 \delta \varphi}{\partial z^2}  - \gamma \frac{b^2_{44}}{c_{44}} M^3_0 \sin \theta_0 \cos^2 \theta_0 \delta \varphi
+ \frac{i\beta_{2}}{2} \cos \theta_0 \left (  e^{-iqz} \frac{\partial s_{+}}{\partial z} -  e^{iqz} \frac{\partial s_{-}}{\partial z} \right ), \\ \label{dphidt}
\sin \theta_0 \frac{\partial \delta \varphi}{\partial t}  &=  - J M_0 \gamma   \frac{\partial^2 \delta \theta }{\partial z^2}
+ \gamma    f(\theta_0)  \delta \theta
+ \frac{\beta_2}{2}  \cos 2 \theta_0 \left (e^{-iqz} \frac{\partial s_{+}}{\partial z}
+  e^{iqz} \frac{\partial s_{-}}{\partial z} \right )
- \beta_4  \sin 2 \theta_0 \frac{\partial s_z}{\partial z},
\end{align}
where a shorthand notation was introduced
\begin{align}
f(\theta_0) = -J q^2 M_0 \cos 2 \theta_0 + H_z \cos \theta_0  + 4 \frac{b^2_{44}}{c_{44}} M^3_0 \sin^2 \theta_0 \cos^2 \theta_0 \nonumber\\
+ 2M_0 u^{(0)}_{xx} (b_{11}-2b_{13}+b_{12}) \cos 2 \theta_0  - 2 \left( b_{33} - b_{31} \right) u^{(0)}_{zz}  M_0 \cos 2 \theta_0,
\end{align}
together with the parameters $\beta_1 = M^2_0 b_{44}/\rho$, $\beta_2 = \gamma M_0 b_{44}$, $\beta_3 = M^2_0 \left( b_{33} - b_{31} \right)/\rho$, $\beta_4 = \gamma M_0 \left( b_{33} - b_{31} \right)$, $v^2_t = c_{44}/\rho$, and $v^2_l = c_{33}/\rho$.
Equations (\ref{dspmdt})-(\ref{dphidt}) can be simplified by transforming into the rotating frame $\tilde{s}_{+} = s_{+} e^{-iqz}$ and $\tilde{s}_{-} = s_{-}e^{iqz}$ that leads to the system with constant coefficients. The dispersion relations for the coupled magnetoelastic waves  can be readily obtained after substituting $e^{ikz-i\omega t}$, that yields at once  the secular equation for the spectrum of coupled magnetoelastic waves
\begin{align} 
&\left[
\left( \omega^2 - \varepsilon_{1k}  \varepsilon_{2k}   \right)
\left(  \omega^2 - v^2_l k^2  \right)
- \beta_3 \beta_4 k^2 \varepsilon_{1k} \sin^2 2 \theta_0
\right]
\left[
\omega^2 - v^2_t \left( k+q \right)^2
\right]
\left[
\omega^2 - v^2_t \left( k-q \right)^2
\right] \nonumber\\
&+ \beta_1 \beta_2  \left( \omega^2 - v^2_l k^2  \right)  \left\{
4 kq \omega^3 \cos \theta_0 \cos 2 \theta_0  -
\left[
\varepsilon_{1k} \cos^2 2 \theta_0 + \varepsilon_{2k} \cos^2  \theta_0
\right]
\left[
\omega^2 \left( k^2 + q^2 \right)  - v^2_t \left( k^2 - q^2 \right)^2
\right]
\right\} \nonumber\\
&- \beta_1 \beta_2 \cos^2 \theta_0 \left\{
\beta_1 \beta_2 \cos^2 2 \theta_0 \left( \omega^2 - v^2_l k^2 \right) \left( k^2 - q^2 \right)^2
+ \beta_3 \beta_4   k^2  \sin^2 2 \theta_0 \left[
\omega^2 \left( k^2 + q^2 \right)  - v^2_t  \left( k^2 - q^2 \right)^2
\right]
\right\}=0,\label{dsxydt}
\end{align}
\end{widetext}where
\begin{align}
\varepsilon_{1k}  &  =\gamma JM_{0}k^{2}+\gamma\frac{b_{44}^{2}}{c_{44}}%
M_{0}^{3}\cos^{2}\theta_{0},\label{ep1}\\
\varepsilon_{2k}  &  =\gamma JM_{0}k^{2}+\gamma f(\theta_{0}). \label{ep2}%
\end{align}
Apparently, the result for a simple spiral is restored for $\theta_{0}=\pi/2$.
In this case, the equation above splits into the dispersion relation for the
longitudinal sound wave, $\omega=v_{l}k$, decoupled from the rest part of the
spectrum for interacting magnetic and transverse sound waves
\begin{align}
\left(  \omega^{2}-\varepsilon_{1k}\varepsilon_{2k}\right)  \left[  \omega
^{2}-v_{t}^{2}\left(  k+q\right)  ^{2}\right]  \left[  \omega^{2}-v_{t}%
^{2}\left(  k-q\right)  ^{2}\right] \nonumber\\
-\beta_{1}\beta_{2}\varepsilon_{1k}\left[  \omega^{2}\left(  k^{2}%
+q^{2}\right)  -v_{t}^{2}\left(  k^{2}-q^{2}\right)  ^{2}\right]  =0.
\end{align}

\subsubsection{Magnetoelastic spectrum in the conical phase}

\begin{figure*}[ptb]
\includegraphics[scale=0.3]{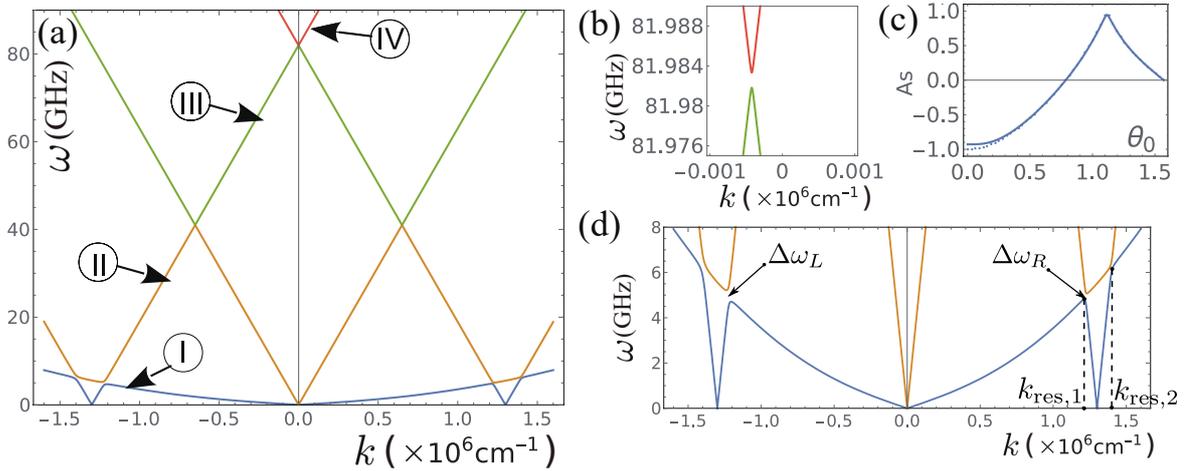}
\caption{(color online) (a) Spectrum of magnetoelastic waves in the conical
phase for $\theta_{0}=\pi/5$ showing four magnetoelastic bands originating from one helimagnon mode and tree acoustic modes -- longitudinal, and left/right-polarized transverse modes. Different energy bands are highlighted by colors. Note a gap opening between each pair of adjacent bands. b) Magnified image of a gap opening between III and IV bands
shifted from $k = 0$ point. c) Band-gap asymmetry, $As = \left(  \Delta\omega_{R} - \Delta\omega_{L} \right)  /\left(\Delta\omega_{R} + \Delta\omega_{L} \right)$, between the left ($\Delta\omega_{L}$) and right ($\Delta\omega_{R}$) gap values centered near $-k_{\mathrm{res},1}$ and $k_{\mathrm{res},1}$ for the hybridized low-energy bands I and II as a function of $\theta_{0}$. d) Low energy sector of the spectrum showing a detailed picture of hybridization between I and II bands.}%
\label{Fig2}%
\end{figure*}

Figure~\ref{Fig2} demonstrates the magnetoelastic spectrum in the conical
phase calculated numerically from Eq.~(\ref{dsxydt}), which shows four
magnetoelastic bands originating from one helimagnon  mode and tree acoustic
modes. The origin of these four bands is intuitively clear -- the lowest
energy mode, I, is a helimagnon-like band except the resonant regions where it
becomes hybridized with right- and left-polarized acoustic bands. Here, we
note a pronounced asymmetry in the degree of hybridization which is discussed
below in detail (see Fig.~\ref{Fig2}~(c)). An important point to note is the
absence of a magnetoelastic gap at $k=0$, the Higgs's effect, owing to the
non-uniform equilibrium strains. The remaining branches II, III, and IV are
acoustic-like bands originating from longitudinal and transverse acoustic
bands hybridized due to the interaction with magnetic excitations. This
interaction generates a gap between each pair of adjacent bands. For example,
a small gap-opening between III and IV bands, which corresponds to the
hybridized left-/right-polarized acoustic bands, is shown in \ref{Fig2}~(b).
Note that the avoided band crossing is shifted from $k=0$, which can be
ascribed to the acoustic activity in the conical phase. In what follows, we
will mainly concentrate on the low-energy part of the spectrum
(Fig.~\ref{Fig2}~(c)), where the magnetization dynamics is coupled to the
elastic subsystem in the most explicit way.

Let us discuss the magnetoelastic resonance between I and II bands. The
momentum points of the first resonance in the vicinity of $\pm q$ (see
Fig.~\ref{Fig2}~(d)) are resulted from the equations
\begin{equation}
v_{t}^{2}\left(  k_{\text{res}}\pm q\right)  ^{2}=\left(  \gamma
JM_{0}\right)  ^{2}k_{\text{res}}^{2}\left(  k_{\text{res}}^{2}+q^{2}\sin
^{2}\theta_{0}\right)  . \label{kres1}%
\end{equation}
By using the values $\gamma=2\pi g\times1.4\,\text{MHz}\cdot\text{Gs}^{-1}$
($g=2$), $q=-0.13\times10^{7}\text{cm}^{-1}$, that corresponds to the period 48
nm, $M_{0}=1649$ Gs, and $JM_{0}^{2}\sim k_{B}T_{c}/a_{||}=0.26\times10^{-6}$
erg/cm, where $T_{c}=127$ K is the Curie-Weiss temperature and $a_{||}%
=6.847\mathring{A}$ is the nearest Cr-Cr distance along the $z$-axis in
{CrNb$_{3}$S$_{6}$},\cite{Mandrus2013} one may find that the pair of resonance
points are given by $k_{\text{res},1}=0.12\cdot10^{7}\text{cm}^{-1}$ and
$k_{\text{res},2}=0.14\cdot10^{7}\text{cm}^{-1}$, where we suppose $\theta_0=\pi/2$. Then the resonance frequency
$\omega_{\text{res}}=\sqrt{c_{44}/\rho}\left\vert k_{\text{res}}+q\right\vert
$ takes the values 5.93 GHz and 7.59 GHz at these points, respectively.

\subsubsection{Band-gap asymmetry in the conical phase}

As anticipated, there is the asymmetry between the left and right gap values,
$\Delta\omega_{L}$ and $\Delta\omega_{R}$, centered near $- k_{\text{res}}$
and $k_{\text{res}}$, respectively, which occurs due to broken parity symmetry
along the $z$ axis in the conical phase. Taking the notation for the
left-hand side of Eq.(\ref{dsxydt}) as $f(\omega)$, the gap in the resonant
point of the frequency $\omega_{\text{res}}$ may be evaluated
\begin{equation}
\label{Gap1}\Delta\omega= 2 \left\{  \left[  \frac{ f^{^{\prime}}%
(\omega_{\text{res}})}{f^{^{\prime\prime}}(\omega_{\text{res}})} \right]  ^{2}
-2 \frac{f(\omega_{\text{res}})}{ f^{^{\prime\prime}}(\omega_{\text{res}})}
\right\}  ^{\frac{1}{2}}.
\end{equation}
The asymmetry between the gaps, defined as $As = \left(  \Delta\omega_{R} -
\Delta\omega_{L} \right)  /\left(  \Delta\omega_{R} + \Delta\omega_{L}
\right)  $ calculated both numerically and with the aid of the formula
(\ref{Gap1}) is shown in Fig.~\ref{Fig2}~(c). It is clear that with decreasing
$\theta_{0}$ the asymmetry gradually increases to some maximum value around
$\theta_{0} = \pi/3$ and drops down afterwards to the minimum value at zero
that corresponds to the forced ferromagnetic state. Mathematically, the
asymmetry in the conical phase results from the $k$-linear term in
(\ref{dsxydt}). It includes the factor $\cos\theta_{0} \cos2 \theta_{0}$ that
reaches the maximum absolute value at $\theta_{0}=0$ and $\cos^{-1}\left(
1/\sqrt{6}\right)  $, and zero at $\pi/4$ and $\pi/2$. This fact explains the
absence of the asymmetry at the last particular points. The symmetry breaking
of the dispersion spectrum admits non-reciprocal elastic wave propagation
controlled by the external magnetic field directed along the chiral axis.

It should be emphasized that the asymmetry indicates involvement of elastic
waves of different polarizations in the hybridization. To illustrate this
fact, let us, at first, have a look at the well known result for the forced
ferromagnetic phase, where only left-polarized transverse acoustic wave
($s_{-} \ne0$), propagating along the magnetic ordering direction, is
hybridized to the magnon band\cite{Akhiezer1968}. This fact is a direct
consequence of the rotations symmetry along the magnetization direction, which
makes polarization of the wave a good quantum number. Since ferromagnetic
magnons are only left-polarized, they are able to couple only to the sound
wave that matches their handedness. 

\begin{figure*}[t]
\begin{center}
\includegraphics[width=15cm]{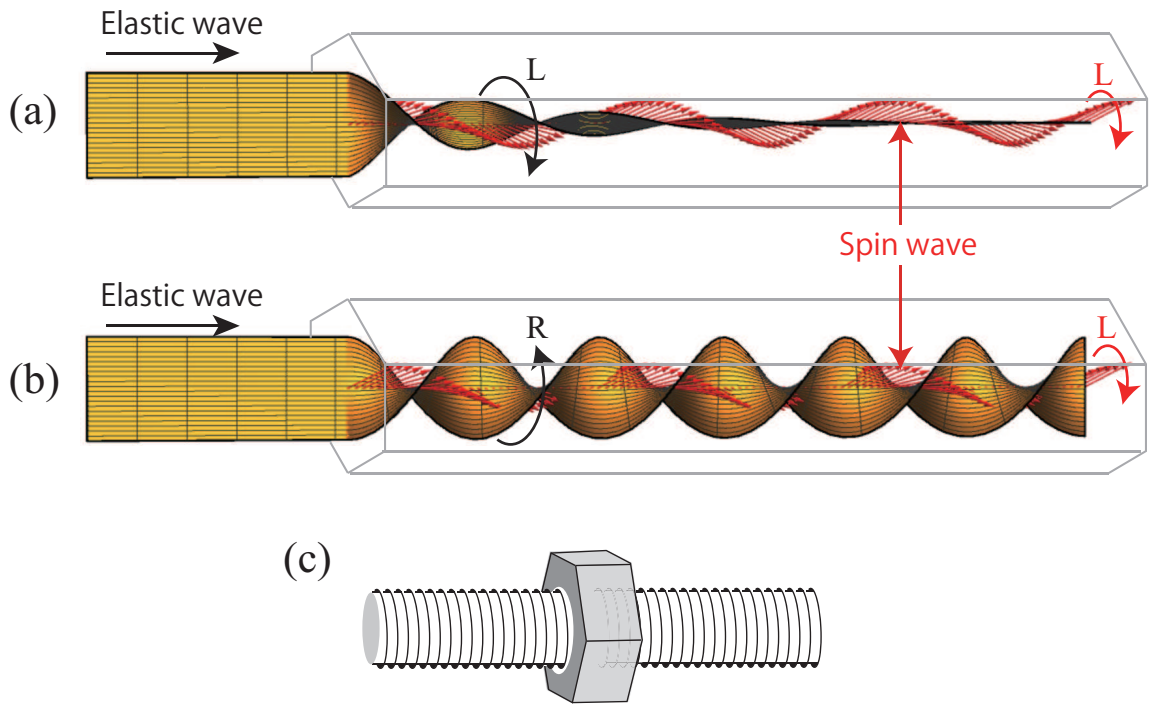}
\end{center}
\caption{(color online) (a) When linearly polarized elastic wave is injected  into the forced-ferromagnetic state along the chiral axis, it is decomposed into left- and right-handed circularly polarized waves.
Then only either of them can resonantly hybridize with  the spin wave which has the definite helicity due to the DM interaction and consequently attenuates. (b) The circularly polarized elastic wave with the chirality opposite to the spin wave can penetrate without attenuation. 
 (c) Mechanical bolt-but analogue of the effect. The screw bolt and the nut correspond to the spin-wave and circularly polarized elastic wave. The nut can couple (hybridize) only when the chirality of the bolt matches the chirality of the nut. This situation is an analogue of the case shown in (a). }%
\label{Fig3}%
\end{figure*}

In Fig.~\ref{Fig3}, we
schematically depict how the left-polarized acoustic wave selectively
hybridize the spin wave in non-reciprocal manner, when the linearly polarized elastic wave is
 injected  into the forced-ferromagnetic state along the chiral axis.
In (a) we show that only either of left- or right-handed circularly polarized counterpart can hybridize with  the spin wave which has the definite helicity due to the DM interaction. In this case, the corresponding counterpart attenuates.  On the other hand, as in Fig. \ref{Fig3} (b) the circularly polarized elastic wave with the chirality opposite to the spin wave can penetrate without attenuation.  This mechanism may be captured through \lq\lq chiral bolt-nut\rq\rq analogue as shown in  Fig. \ref{Fig3} (c).
In the case of conical phase with $\theta_0\neq\pi/2$, left- and right-handed spin waves are mixed and consequently, the linearly polarized elastic wave are decomopsed into left- and right-handed circularly polarized counterparts depending on the magnitude of $\theta_0$.

\begin{figure}[t]
\begin{center}
\includegraphics[width=7cm]{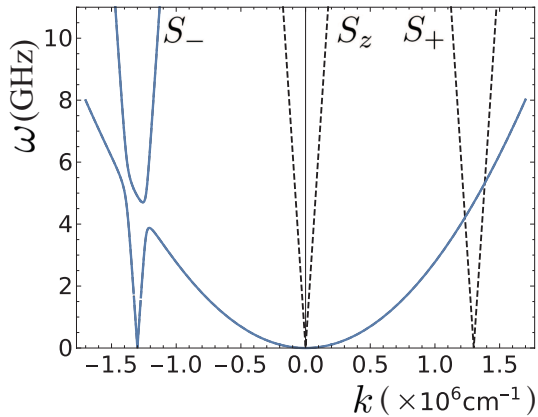}
\end{center}
\caption{Spectrum of magnetoelastic waves at $\theta_{0}=0$, where the conical
phase collapses to the forced ferromagnetic phase, calculated in the two-wave
approximation (see Eq.~\eqref{2wap}), which merges perfectly with the exact
result given by Eq.~\eqref{dsxydt}. The dashed lines show the longitudinal
and right-polarized acoustic waves decoupled from magnetic excitations. The
band gap on the left hand side shows remaining hybridization between the
parabolic ferromagnetic magnon spectrum and left-polarized acoustic wave.}%
\label{Fig4}%
\end{figure}

The same argument is applicable to the conical phase at $H^{z} = H_{c}^{z}$
($\theta_{0} = 0$) where rotation symmetry is restored, see Fig.~\ref{Fig4}.
However, for $H^{z} < H_{c}^{z}$ the finite component of the magnetization
appears perpendicular to the chiral axis, which breaks the rotation symmetry
giving rise to the direct hybridization between right-polarized ($s_{+} \ne0$)
acoustic band and the helimagnon band. Therefore, the asymmetry factor, $As$,
can be related to the difference between the contributions from the left- and
right-polarized acoustic waves to the hybridization.

To summarize this section, we would like to note that the measurements of the
band gaps in the spectrum of magnetoelastic excitations can be useful for
experimental estimation of magnetoelastic constants. As an example, we
demonstrate how the constant $b_{44}$ responsible for hybridization between I
and II bands may be determined from the experimental value for the band gap at
$\theta_{0}=0$ of the phase transition from the conical phase to the induced
ferromagnetic phase (see Fig.~\ref{Fig4}).

The choice of this specific point is motivated by the absence of contribution
of another magnetoelastic constants to the gap value. For illustration, we
restrict our analysis by two-wave approximation,\cite{two-wave}  where coupling of the
amplitudes $\tilde{s}_{-}(k)$, $\delta\varphi(k)$ and $\delta\theta(k)$ is only
retained in the vicinity of the momentum $q$. Then the frequencies may be
found from determinant of the matrix
\begin{equation}
\label{2wap}\left(
\begin{array}
[c]{ccc}%
\omega^{2} - v^{2}_{t} (k-q)^{2} & \beta_{1} (k-q) & i \beta_{1} (k-q)\\
\frac{1}{2}\beta_{2} (k-q) & - \varepsilon_{1k} & i\omega\\
\frac{i}{2} \beta_{2}(k-q) & i \omega & \varepsilon_{2k}%
\end{array}
\right)  .
\end{equation}
It can be observed from Eqs.(\ref{ep1},\ref{ep2}) that
\begin{equation}
\varepsilon_{1k}=\varepsilon_{2k} = \gamma J M_{0} k^{2} + \frac{\beta_{1}
\beta_{2}}{v^{2}_{t}},
\end{equation}
and, as a consequence, only the constant $b_{44}$ controls interaction between
the magnetic and the elastic subsystems.

Straightforward calculation results in the dispersion relation (see Fig.
\ref{Fig4})
\begin{equation}
\left(  \omega^{2} - v^{2}_{t} (k-q)^{2} \right)  \left(  \omega- \gamma J
M_{0} k^{2} \right)  = \beta_{1} \beta_{2} \frac{\omega^{2}}{v^{2}_{t}}.
\end{equation}
By adapting the resonance condition in Eq.~\eqref{kres1} and using the
expansion $\omega= \omega_{0} + \delta\omega$, where $\omega_{0} = \gamma J
M_{0} k^{2}_{\text{res}}$ is the frequency of the resonance, we find
eventually the gap value
\begin{equation}
\delta\omega\approx\gamma M^{2}_{0} b_{44} \sqrt{\frac{ J k_{\text{res}}^{2}%
}{2c_{44}}}.
\end{equation}
Consequently, the non-transmission band in the spectrum of the coupled
oscillations enables a convenient way to find the magnetoelastic constant
$b_{44}$ associated with torsion deformations around the $z$-axis.

\subsection{Soliton lattice phase}

In this section, we consider the case when the static magnetic field is
applied perpendicular to the chiral axis. For $H^{x} < H_{c}^{x}$, the
magnetic chiral soliton lattice phase is realized, which is characterized by
the following spatial dependence of the equilibrium background magnetization,
$M_{0\pm}(z) = M_{0}e^{\pm i\varphi_{0}(z)}$, and $M_{0z} = 0$, where
$\varphi_{0}(z)$ is given by Eq.~\eqref{SLsol}. At $H^{x} = 0$, the Taylor
series of $\varphi_{0}$ have only one term, $qz$, which corresponds the simple
spiral with one harmonic. For any nonzero $H^{x}$, Jacobi's amplitude function
in $\varphi_{0}(z)$ has nontrivial power series giving origin to the
multiharmonic nature of the resulting soliton lattice.

In order to obtain the spectrum of magnetoelastic excitations for the soliton
lattice phase, we expand the total magnetization up to the linear order in
fluctuations, $\delta m_{\pm}(z,t) = \pm i M_{0} e^{\pm i \varphi_{0} (z)}
\delta\varphi(z,t)$ and $\delta m_{z} (z,t) = - M_{0} \delta\theta(z,t)$. The
dynamical equations can be found straightforwardly from Eq.~(\ref{EM1},
\ref{EM2}) by linearizing them in $\delta\varphi$ and $\delta\theta$, which
gives the following expressions after some algebra
\begin{align}
\frac{\partial^{2} s_{\pm}}{\partial t^{2}}  &  = v^{2}_{t}
\frac{\partial^{2} s_{\pm}}{\partial z^{2}} - \beta_{1} \frac{\partial
}{\partial z} \left(  e^{\pm i\varphi_{0}} \delta\theta\right)  ,\label{PDESL1}\\
\frac{\partial\delta\theta}{\partial t}  &  = -JM_{0} \gamma\Lame \delta
\varphi,\\
\frac{\partial\delta\varphi}{\partial t}  &  = JM_{0} \gamma\left[  \Lame -
\left(  \frac{d \varphi_{0}}{d z} - q \right)  ^{2} \right]  \delta\theta+
\gamma f \left(  \frac{\pi}{2}\right)  \delta\theta\nonumber\\
&  - \frac{\beta_{2}}{2} \left(  e^{-i\varphi_{0}} \frac{\partial s_{+}%
}{\partial z} + e^{i\varphi_{0}} \frac{\partial s_{-}}{\partial z} \right)  \label{PDESL4},
\end{align}
where $\Lame = -\partial_{z}^{2} + m^{2}\cos\varphi_{0}$ denotes the Lam\'{e}
operator, and the sine-Gordon equation, providing the phase modulation in the
soliton lattice, $\partial^{2}_{z} \varphi_{0}=m^{2} \sin\varphi_{0}$, was
accounted for. The equation of motion for $s_{z}$ is totally decoupled from
these equations and corresponds to the acoustic band with the trivial
dispersion relation $\omega= v_{l}k$.

It is natural to assume that some generalized Fourier series for $
s_{\pm}$, $\delta\theta$, and $\delta\varphi$ in terms of the Lam\'e
operator's eigenfunction can provide the solution of the eigenvalue problem
when the magnetoelastic coupling is fairly small. However, in realizing this
approach one faces with a problem, since it turns out that, in practice, it is
not possible to treat this infinite series as being explicitly controlled by
any small parameter whatsoever.

To tackle this problem, let us note the case for the conical phase, where the
gauge transformation for $s_{\pm}$ was applied to remove the periodic
terms in the equations of motion, which appeared owing to the basic harmonics,
$\sin\theta_{0} e^{\pm i qz}$, of the underlying magnetic structure.
Unfortunately, this special trick cannot be directly implemented for
Eqs.~(\ref{PDESL1})-(\ref{PDESL4}) because of the multi-harmonic character of
the soliton lattice phase. Nevertheless, we found that the expansion of the
periodic terms with respect to the small parameter $\kappa^{2}$, which is
controlled by $H^{x}$, with subsequent Fourier transformation of the dynamical
equations turns out to be effective.

Indeed, the coefficients on the right hand side of Eqs.~(\ref{PDESL1}%
)--(\ref{PDESL4}) can be expanded in power series of $\kappa$
\begin{align}
\cos\varphi_{0}  &  = \frac{\kappa^{2}}{8} - \cos qz -
\frac{\kappa^{2}}{8} \cos2qz + \mathcal{O}(\kappa^{4}),\label{Exp1}\\
e^{\pm i \varphi_{0}}  &  = \frac{\kappa^{2}}{8} - e^{\pm iqz} - \frac
{\kappa^{2}}{8} e^{\pm2iqz} + \mathcal{O}(\kappa^{4}),\\
\frac{d \varphi_{0}}{dz}  &  = q + \frac{\kappa^{2}}{4} q \cos qz +
\mathcal{O}(\kappa^{4})\label{Exp3}
\end{align}
where $\kappa$ is determined by applied magnetic field.

One particular advantage of the present formulation is evident for small and
intermediate magnetic fields, when $H^{x}$ is far below $H_{c}^{x}$; because
these expansions involve the small factor $\kappa^{2}$, the series can be
terminated at low order. The method is also sufficiently simple algebraically
to enable us to obtain a magnetoelastic spectrum in the soliton lattice phase
with a given accuracy.

Inserting the expansions (\ref{Exp1})-(\ref{Exp3}) into the system
(\ref{PDESL1}) -(\ref{PDESL4}) and holding terms up to the $\kappa^{2}$ order,
we get \begin{widetext}
\begin{align}  \label{sleq1}
\left( \omega^2 - v_t^2 k^2 \right) s_{\pm} (k,\omega) &=   ik \beta_1 \left [\frac{\kappa^2}{8} \delta \theta (k,\omega) - \delta \theta (k \mp q,\omega)  -  \frac{\kappa^2}{8} \delta \theta (k \mp 2q,\omega)\right ] ,
\\
\label{sleq3}
- i \omega \delta \theta(k,\omega) &= - JM_0 \gamma k^2 \left[ \delta \varphi (k,\omega) - \frac{q^2 \kappa^2}{8k^{2}} \left(  \delta \varphi (k+q,\omega) + \delta \varphi (k-q,\omega) \right)\right ],
\\ \nonumber
- i\omega \delta \varphi (k,\omega) &=  JM_{0}\gamma k^{2}\left[ \delta \theta (k,\omega) - \frac{ q^{2} \kappa^2}{8k^{2}} \left(  \delta \theta (k+q,\omega) +  \delta \theta (k-q,\omega) \right) \right]  + \gamma f \left( \frac{\pi}{2} \right)\delta \theta (k,\omega)
\\ \nonumber
&-\frac{i\beta_{2}\kappa^2}{16}\left[  k s_{+}(k,\omega) + k s_{-}(k,\omega) - (k+2q) s_{+}(k+2q,\omega) - (k-2q) s_{-}(k-2q,\omega) \right]
\\
\label{sleq4}
&+\frac{i\beta_{2}}{2}\left[(k+q) s_{+}(k+q,\omega) + (k-q) s_{-}(k-q,\omega) \right ] .
\end{align}
\end{widetext}

\begin{figure*}[t]
\begin{center}
\includegraphics[width=.75\textwidth]{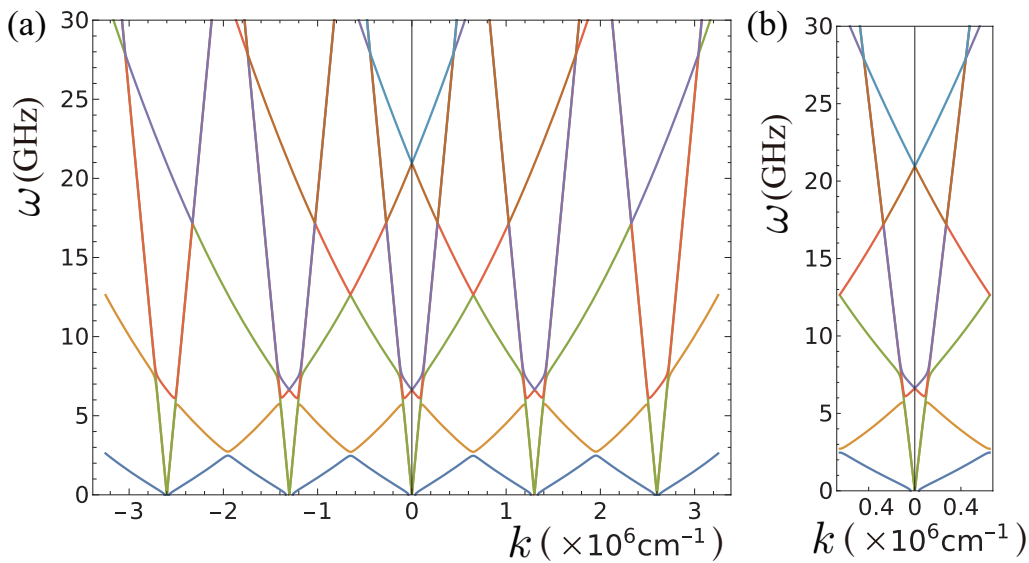}
\end{center}
\caption{(color online)Spectrum of magnetoelastic waves in the soliton lattice
phase in the extended (a) and reduced (b) zone schemes. The colors indicate
different excitation bands separated from each other by hybridization gaps.}%
\label{Fig5}%
\end{figure*}

To obtain a closed set of dynamical equations, we supplemented
Eqs.~(\ref{sleq1})--(\ref{sleq4}) by similar equations of motion for higher
order harmonic amplitudes keeping only the terms with $k \pm q$ and $k \pm2q$.
The resulting set of twenty coupled equations was solved numerically to obtain
magnetoelastic band structure shown in Fig.~\ref{Fig5}.

The resulting band structure in Fig.~\ref{Fig5} can be qualitatively
understood if we note that the periodic nature of the magnetic soliton lattice
gives origin to the magnetic Brillouin zone determined by the soliton lattice
period and controlled by external magnetic field. Magnetic excitation can
directly feel this periodic background which naturally results into the
helimagnon Bloch bands, where different branches are separated from each other
due to the Bragg's reflection from periodic potential of the underlying
magnetic superlattice. These helimagnon Bloch bands hybridize with acoustic
bands due to the magnetoelastic coupling resulting into the energy spectrum
shown in Fig.~\ref{Fig5}~(b).

To gain further insight concerning the excitation spectrum, it may be useful
to decompose the background magnetization of the soliton lattice into the
harmonic series \cite{KO2015}
\begin{align}
\frac{M_{x0}}{M_{0}}  &  = \frac{2(K-E)}{\kappa^{2} K} -1 - \frac{\pi^{2}%
}{\kappa^{2} K^{2}} \sum_{n \not = 0} \frac{n e^{inGz}}{\sinh\left(  n
\pi\frac{K^{\prime}}{K} \right)  } ,\\
\frac{M_{y0}}{M_{0}}  &  = \frac{i \pi^{2}}{\kappa^{2} K^{2}} \sum_{n} \frac{n
e^{inGz}}{\cosh\left(  n \pi\frac{K^{\prime}}{K} \right)  },
\end{align}
where
\begin{equation}
G =\pi^{2} q/(4KE) = q \left[  1 - \frac{\kappa^{4}}{32} + \mathcal{O} \left(
\kappa^{6} \right)  \right]
\end{equation}
is the wave vector of the soliton lattice, and $K$ ($K^{\prime}$) denotes the
first order elliptic integral with the modulus $\kappa$ ($\kappa^{\prime
2})^{1/2}$). In contrast to the conical phase, the additional contributions
$e^{i n G z}$, $|n| \geq2$, appear in the spatial distribution of the
nonuniform magnetic background along with the basic ones, $e^{\pm i G z}$.

Inspection of Fig.~\ref{Fig5}~(a) indicates that we can assign different
coordinate systems related to each harmonic, where the points $nq$ are used as
the coordinate system origin, and, as a consequence, the excitation branches
of the elastic excitations are replicated. Similarly to the simple spiral, the
resonance at $k_{\text{res},\alpha}$ ($\alpha=1,2$) points near the $nq$
values occurs, which is determined by the following condition
\begin{equation}
\label{kres}v^{2}_{t} \left(  k_{\text{res},\alpha} \pm n q \right)  ^{2} =
\varepsilon_{1k} \varepsilon_{2k}%
\end{equation}
giving resonant frequencies $\omega^{(n)}_{\text{res},\alpha} =\sqrt
{c_{44}/\rho} |k_{\text{res},\alpha} \pm nq|$. By neglecting the
magnetoelastic contributions to the energies $\varepsilon_{1k,2k}$, we recover
the result of Eq.~(\ref{kres1}).

Proceeding similarly to the analysis of the conical phase, one may observe
that the first gap in the excitation spectrum in the vicinity of $k=q$
originates from hybridization of the amplitudes $s_{-}(k-q)$, $\delta
\theta(k)$ and $\delta\varphi(k)$. The system (\ref{sleq1},\ref{sleq3}%
,\ref{sleq4}) lends support to the coupling
\begin{align}
\label{sleq3g1}\left[  \omega^{2} \! - v^{2}_{t} (k-q)^{2} \right]  \! s_{-}
(k-q) + i\beta_{1} (k-q) \delta\theta(k) = 0 ,\\
i \omega\delta\theta(k) - \varepsilon_{1k} \delta\varphi(k) =0,\\
i\omega\delta\varphi(k) + \varepsilon_{2k} 
\delta\theta(k) + \frac{i
\beta_{2}}{2} (k-q) s_{-}(k-q) = 0
\end{align}
which brings about the result for the first hybridization gap between the
magnetic and acoustic band
\begin{equation}
\label{GapSplit0}\left.  \Delta\omega\right|  _{k= q} \approx\gamma M^{2}_{0}
b_{44} \sqrt{\frac{Jq^{2}}{2c_{44}}}.
\end{equation}
The extension of this approach to calculation of the second order gap seems
obvious. Apparently, keeping only the amplitudes $s_{-}(k-2q)$, $\delta
\theta(k)$ and $\delta\varphi(k)$ in Eqs~(\ref{sleq1})--(\ref{sleq4}), one
finds the gap near the resonant point $k=2q$
\begin{equation}
\label{GapSplit2}\left.  \Delta\omega\right|  _{k= 2 q} \approx\gamma
M^{2}_{0} b_{44} \frac{\kappa^{2}}{4} \sqrt{\frac{Jq^{2}}{2c_{44}}} =
\frac{\kappa^{2}}{4} \left.  \Delta\omega\right|  _{k= q}.
\end{equation}
It may be further proved that the width of the $n$-th gap decreases
exponentially, $\left.  \Delta\omega\right|  _{k= n q} \sim\kappa^{2n-2}$,
similar to the result for spin-wave spectrum of the relativistic spiral
\cite{Izyumov1987}.

Apart from hybridization between the spin and elastic waves, there is a pure
magnetic band gap originating from the Bragg's reflection of the helimagnons
from the periodic potential of the soliton lattice. It can be regarded as the
splitting $\Delta\omega_{\rm{sp}}= a^{3}M_{0}H^{x}/(2\hbar)$ between the acoustic and
optic branches of spin fluctuations at the boundary of the magnetic Brillouine
zone \cite{KO2015}, which is visible in Fig.~\ref{Fig5}~(a) as lifted
degeneracies at the points $nG/2 \approx nq/2$. In contrast to
Eq.~(\ref{GapSplit0}), the magnetic gap is directly controlled by the magnetic
field rather than a strength of the magnetoelastic coupling.

\section{Discussions}

A salient peculiarity of the conical phase is the conspicuous asymmetry
between the left and right band-gaps in the spectrum of the coupled
magnetoelastic waves. In practice, it is the phonon mode that of major
importance after hybridization, because the elastic stiffness is measured
experimentally at different external magnetic fields as the ultrasonic
response. The tunable non-reciprocity governed by the magnetic field is
potentially applicable in the construction of ultrasound devices using chiral helimagnets.

In contrast to the conical magnetic structure, time-reversal symmetry for
elastic wave propagation is kept for the soliton lattice and for the simple
spiral, particularly. Another notable difference in comparison with the
conical phase, the magnetoelastic resonance in the soliton lattice has the
multi-resonance behavior. This result confirms the intuitive expectation that
the resonance occurs whenever the wave vector of a spreading elastic wave
matches a modulation of the non-uniform magnetic background. In contrast to
the conical magnetic structure, the soliton lattice consists of higher-order
harmonics indexed by integer, and each of the components contributes to the
resonance separately. We emphasize that an assessment of the hybridization
constant $b_{44}$ at the point, where the conical phase is collapsed in favor
of the forced ferromagnetic phase, may successfully be combined with
measurements of multiresonance ultrasound absorption in the soliton lattice.
The scheme provides a promising tool for an experimental probe of the soliton
lattice phase. Regarding potential applications of the theory, it is useful to
highlight that while lattice and elastic properties of MnSi and related
compounds are well known \cite{Lamago2010,Petrova2011}, there remains a
considerable need for experimental information on the phonon dispersion and
the phonon density of states in {CrNb$_{3}$S$_{6}$}.

While our treatment is designed for crystals of hexagonal symmetry it
nonetheless provides the framework for studies of magnetoelastic effects in
chiral helimagnets of other crystal classes. For example, the tetragonal
insulating materials CuB${}_{2}$O${}_{4}$ \cite{Roessli2001} and Ba${}_{2}%
$CuGe${}_{2}$O${}_{7}$ \cite{Zheludev1998}, the trigonal metallic compound
Yb(Ni${}_{1-x}$Cu${}_{x}$)${}_{3}$Al${}_{9}$ \cite{Matsumura2017} may be
named, where ample evidences for the formation of a chiral magnetic soliton
lattice state, an anticipated outcome of a monoaxial chiral helimagnet, were reported.

Some limitations of our analysis should be mentioned. In the equilibrium
configuration $\boldsymbol{M}_{0}$, the magnetoelastic terms were discarded. These effects may be described by the double sine-Gordon
model, also known as the sine-Gordon model with crystalline anisotropy of the
second order \cite{Izyumov1984}. This specific issue will be addressed in
future work. Here, it is worth noting that the enhanced anisotropic change in
shape both for skyrmion lattice and individual skyrmions was revealed in FeGe
by Lorentz transmission electron microscopy under uniaxial tensile stress
deformation. It was ascribed to the strain-induced anisotropic modulation of
DM interaction \cite{Shibata2015}. On the contrary, the stress-driven
topological phase transition in MnSi from the skyrmion lattice phase to the
conical phase was interpreted by strain-induced magnetic anisotropy on the
basis of the Ginzburg-Landau phenomenology with an account of magnetoelastic
contribution to the free energy \cite{Nii2015}.

Another difficulty of possible application of the work may arise owing to the
magneto-elastic correlations in {CrNb$_{3}$S$_{6}$} \cite{Mito2015}. The
diffuse scattering measurements of the crystal structure of {CrNb$_{3}$S$_{6}%
$} demonstrate that there is a bias towards a disorder in the Cr sublattice
\cite{Chernyshov2015}. It is suggesting that the disorder occurs due to
clustering of Cr ions in hexagonal fragments within the layers. It was found
that such a specific correlated disorder strongly affects the magnetic
ordering temperature. A follow up work designed to evaluate an interplay
between the correlated disorder and magnetic properties would be useful.

Measurements on thin films of {CrNb$_{3}$S$_{6}$} showed that the chiral
soliton lattice exhibits interesting phenomena due to confinement from the
presence of magnetic domains extended for approximately 1 $\mu$m in helix
direction \cite{Togawa2015,Wang2017}. An important question for future studies
is to determine an effect of the domain structure on the ultrasound wave
propagation. We believe that our theoretical analysis may serve as an
appropriate starting point to touch on these issues.

\section{Conclusions}

In summary, we have investigated the spectrum of coupled magnetoelastic waves
propagating along the helicoidal axis in crystals of hexagonal symmetry having
spiral magnetic order due to DM interaction. Based on the
example of spin and elastic waves we elucidate how torsion deformations are
related with spin chirality. We clarified peculiar nature of magnetoelastic
resonance for particular phases of the monoaxial chiral axis: the conical
phase and the soliton lattice phase. To the best of our knowledge, an effect
of magnetoelastic coupling for the latter one has not been studied before. 

So far some kinds of multiresonance phenomena associated with the soliton lattice have been predicted, including an appearance of higher-order satellites in the neutron diffraction patterns
\cite{Izyumov1984,KO2015}, a spike-like behavior of magnetoresistance originated
from scattering of electrons by the magnetic superlattice by the chiral solitons
\cite{KPO2011,Okamura2018}, and multiple spin resonance of the chiral soliton
lattice \cite{KO2009}.  We expect the present study on magneto-elastic coupling may expand the scope of these multi-resonance or scattering phenomena. In particular, we show that the non-reciprocal spin wave around the forced-ferromagnetic state has potential capability to convert the linearly polarized elastic wave to circularly polarized one with the chirality (helicity) opposite to the spin wave chirality.

\bigskip

\begin{acknowledgments}
Special thanks are due to N. Baranov for very informative discussions at various stages. 
We also thank Masaki Mito and Yoshihiko Togawa for enlightening discussions on magnetoelastic problem over the years. The work was supported by the Government of the Russian Federation Program
02.A03.21.0006.
This work was also supported by a Grant-in-Aid for Scientific
Research (B) (No. 17H02923) and (S) (No. 25220803) from the MEXT of the
Japanese Government, JSPS Bilateral Joint Research Projects (JSPS-FBR), and the JSPS Core-to-Core Program, A. Advanced Research Networks. I.P.
acknowledges financial support by Ministry of Education and Science of the Russian Federation, Grant No. MK-1731.2018.2. A.A.T. and I.P. are also supported by Russian Foundation for Basic Research (RFBR), Grant 18-32-00769(mol\_a). A.S.O. acknowledge funding by the RFBR, Grant 17-52-50013, and  the Foundation for the Advancement to Theoretical Physics and Mathematics BASIS Grant No. 17-11-107.
\end{acknowledgments}


\begin{thebibliography}{99}                                                                                               %


\bibitem {Fawcett1970}E. Fawcett, J.P. Maita and J.H. Wernick, Int. J. Magn.
\textbf{1}, 29 (1970).

\bibitem {Matsunaga1982}M. Matsunaga, Y. Ishikawa and T. Nakajima, J. Phys.
Soc. Japan \textbf{51}, 1153 (1982).

\bibitem {Valkovskiy2016}G.A. Valkovskiy, E.V. Altynbaev, M.D. Kuchugura, E.G.
Yashina, A.S. Sukhanov, V.A. Dyadkin, A.V. Tsvyashchenko, V.A. Sidorov, L.N.
Fomicheva, E. Bykova, S.V. Ovsyannikov, D.Yu. Chernyshov and S.V. Grigoriev,
J. Phys.: Condens. Matter \textbf{28}, 375401 (2016).

\bibitem {Makarova2012}O. L. Makarova, A. V. Tsvyashchenko, G. Andre, F.
Porcher, L. N. Fomicheva, N. Rey, and I. Mirebeau, Phys. Rev. B \textbf{85},
205205 (2012).

\bibitem {Dyadkin2014}V. Dyadkin, S. Grigoriev, S.V. Ovsyannikov, E. Bykova,
L. Dubrovinsky, A. Tsvyashchenko, L.N. Fomicheva, and D. Chernyshov, Acta
Crystallogr. B Struct. Sci. Cryst. Eng. Mater. \textbf{70} 676, (2014).

\bibitem {Martin2016}N. Martin, M. Deutsch, J.P. Iti\'e, J.-P. Rueff, U.K.
R\"ossler, K. Koepernik, L.N. Fomicheva, A.V. Tsvyashchenko, and I. Mirebeau,
Phys. Rev. B \textbf{93}, 214404 (2016).

\bibitem {Plumer1982}M.L. Plumer and M.B. Walker, J. Phys. C: Solid State
Phys. \textbf{15}, 7181 (1982).

\bibitem {Plumer1984}M.L. Plumer, J. Phys. C: Solid State Phys. \textbf{17},
4663 (1984).

\bibitem {Maleyev2009}S.V. Maleyev, J. Phys.: Condens. Matter \textbf{21},
146001 (2009).

\bibitem {Petrova2009}A.E. Petrova, S.M. Stishov, J. Phys.: Condens. Matter
\textbf{21}, 1960001 (2009).

\bibitem {Petrova2016}A.E. Petrova, S.M. Stishov, Phys. Rev. B \textbf{91},
214402 (2016).

\bibitem {Togawa2012}Y. Togawa, T. Koyama, K. Takayanagi, S. Mori, Y. Kousaka,
J. Akimitsu, S. Nishihara, K. Inoue, A. S. Ovchinnikov, and J. Kishine, Phys.
Rev. Lett. \textbf{108}, 107202 (2012).

\bibitem {Binz2009}S. M\"uhlbauer, B. Binz, F. Jonietz, C. Pfleiderer, A. Rosch, A. Neubauer, R. Georgii, and P. B\"oni, Science
\textbf{323}, 915 (2009).

\bibitem {Yu2010}X. Z. Yu, Y. Onose, N. Kanazawa, J. H. Park, J. H. Han, Y.
Matsui, N. Nagaosa, and Y. Tokura, Nature (London) \textbf{465}, 901 (2010).

\bibitem {Adams2010}W. M\"unzer, A. Neubauer, T. Adams, S. M\"uhlbauer, C.
Franz, F. Jonietz, R. Georgii, P. B\"oni, B. Pedersen, M. Schmidt, A. Rosch,
and C. Pfleiderer, Phys. Rev. B \textbf{81}, 041203(R) (2010).


\bibitem {Seki2012}S. Seki, X. Z. Yu, S. Ishiwata, and Y. Tokura, Science
\textbf{336}, 198 (2012).



\bibitem {Nii2015}Y. Nii, T. Nakajima1, A. Kikkawa, Y. Yamasaki, K. Ohishi, J.
Suzuki, Y. Taguchi, T. Arima, Y. Tokura and Y. Iwasa, Nat.Commun. \textbf{6},
8539 (2015).

\bibitem {Kittel1958}C. Kittel, Phys. Rev. \textbf{110}, 836 (1958).

\bibitem {Stefanovski1969}V.G. Bar'yahtar and E.P. Stefanovski, Fiz. Tverd.
Tela \textbf{11}, 1946 (1969).

\bibitem {Vlasov1973}K.B. Vlasov, V.G. Bar'yahtar and E.P. Stefanovski, Fiz.
Tverd. Tela \textbf{15}, 3656 (1973).

\bibitem {Turov1983}E.A. Turov and V.G. Shavrov, Sov. Phys. Usp. \textbf{26},
593 (1983).

\bibitem {Shavrov1989}V.D. Buchel'nikov, V.G. Shavrov, Sov. Phys. Solid State
\textbf{31}, 23 (1989).

\bibitem {Bychkov1990}V.D. Buchel'nikov, I.V. Bychkov and V.G. Shavrov, Fiz.
Met. Metalloved. \textbf{11}, 12 (1990).

\bibitem {Vittoria2015}C. Vittoria, Phys. Rev. B \textbf{92}, 064407 (2015).

\bibitem {Iguchi2015}{Y. Iguchi, S. Uemura, K. Ueno, Y. Onose, Phys.
Rev. B \textbf{92} (2015)184419. }



\bibitem {Seki2015}{S. Seki, Y. Okamura, K. Kondou, K. Shibata, M. Kubota, R. Takagi, F. Kagawa, M. Kawasaki, G. Tatara, Y. Otani, and Y. Tokura, Phys. Rev. B \textbf{93},
235131 (2016). }

\bibitem {Zhang2017}X.-X. Zhang and N. Nagaosa, New J. Phys. \textbf{19},
043012 (2017).


\bibitem {Kanazawa2016}N. Kanazawa, Y. Nii, X.-X. Zhang, A.S. Mishchenko, G.
De Filippis, F. Kagawa, Y. Iwasa, N. Nagaosa and Y. Tokura, Nat.Commun.
\textbf{7}, 11622 (2016).

\bibitem {Hu2017}Y. Hu and B. Wang, New J. Phys. \textbf{19}, 123002 (2017).

\bibitem {Rosch2016}A. Rosch, Nat. Mater. \textbf{15}, 1231 (2016).

\bibitem {Shavrov1993}V.D. Buchel'nikov, I.V. Bychkov and V.G. Shavrov, J.
Magn. Magn. Mater. \textbf{118}, 169 (1993).


\bibitem {Mason1954}W.P. Mason, Phys. Rev. \textbf{96}, 302 (1954).


\bibitem {Comstock1963}R.L. Comstock and B.A. Auld, J. Appl. Phys.
\textbf{34}, 1461 (1963).




\bibitem {Mandrus2013}N.J. Ghimire, M.A. McGuire, D.S. Parker, B. Sipos, S.
Tang, J.-Q. Yan, B.C. Sales, and D. Mandrus, Phys. Rev. B \textbf{87}, 104403 (2013).


\bibitem {Gaillac2016}R. Gaillac, P. Pullumbi and F.-X. Coudert, J. Phys.
Condens. Matter \textbf{28}, 275201 (2016).


\bibitem {Akhiezer1968}A. I. Akhiezer, V. G. Baryakhtar, and S.
V. Peletminskii, in Spin Waves, edited by S. Doniach (North-Holland Publishing
Co., Amsterdam, 1968).

\bibitem{two-wave} Two-wave approximation used here is analogous to the case of magnetic resonance problem.
See, for example, C.~P.~Slichter, \lq\lq
Principles of Magnetic Resonance,\rq\rq
(Springer, 1990).  The role of AC magnetic field in the magnetic resonance problem is played by the elastic wave in the present case.


\bibitem {Higgs1964}P.H. Higgs, Phys. Lett. \textbf{12}, 132 (1964).

\bibitem {Shavrov1994}V.D. Buchel'nikov, I.V. Bychkov, V.G. Shavrov, JETP
\textbf{78}, 398 (1994).

\bibitem {Chandra1994}D. Chandrasekharaiah and L. Debnath, \textit{Continuum
Mechanics} (Academic Press, San Diego, 1994).

\bibitem {Fedorov1997}V.I. Fedorov, A.G. Gukasov, V. Kozlov, S.V. Maleyev,
V.P. Plakhty, I.A. Zobkalo, Phys. Lett. A \textbf{224}, 372 (1997).

\bibitem {Grigoriev2009}S. V. Grigoriev, D. Chernyshov, V. A. Dyadkin, V.
Dmitriev, S. V. Maleyev, E. V. Moskvin, D. Menzel, J. Schoenes, and H.
Eckerlebe, Phys. Rev. Lett. \textbf{102}, 037204 (2009).

\bibitem {Grigoriev2010}S.V. Grigoriev, D. Chernyshov, V.A. Dyadkin, V.
Dmitriev, E.V. Moskvin, D. Lamago, Th. Wolf, D. Menzel, J. Schoenes, S.V.
Maleyev, and H. Eckerlebe, Phys. Rev. B \textbf{81}, 012408 (2010).

\bibitem {KO2015}J. Kishine and A. S. Ovchinnikov, Solid State Phys. 66, 1 (2015).

\bibitem {Izyumov1987}Yu.A. Izyumov, \textit{Diffraction of Neutrons on
Long-Periodic Structures} [in Russian] (Energoatomizdat, Moscow, 1987).

\bibitem {Lamago2010}D. Lamago, E. S. Clementyev, A. S. Ivanov, R. Heid, J.-M.
Mignot, A. E. Petrova, and P. A. Alekseev, Phys. Rev. B \textbf{82}, 144307 (2010).

\bibitem {Petrova2011}A.E. Petrova, V.N. Krasnorussky, W.M. Yuhasz, T.A.
Lograsso and S.M. Stishov, J. Phys.: Conf. Ser. \textbf{273}, 012056 (2011).

\bibitem {Izyumov1984}Y.A. Izyumov, Sov. Phys. Usp. \textbf{27}, 845 (1984).

\bibitem {KPO2011}J. Kishine, I. V. Proskurin, and A. S. Ovchinnikov, Phys.
Rev. Lett. \textbf{107}, 017205 (2011).

\bibitem {Okamura2018}S. Okamura, Y. Kato and Y. Motome, J. Phys. Soc. Jpn (in press).

\bibitem {KO2009}J. Kishine and A. S. Ovchinnikov, Phys. Rev. B \textbf{79},
220405(R) (2009).

\bibitem {Roessli2001}B. Roessli, J. Shefer, G.A. Petrakovskii, B. Ouladdiaf,
M. Boehm, U. Staub, A. Vorotinov, L. Bezmaternikh, Phys. Rev. Lett.
\textbf{86}, 1885 (2001).

\bibitem {Zheludev1998}A. Zheludev, S. Maslov, G. Shirane, Y. Sasago, N. Koide
and K. Uchinokura, Phys. Rev. B \textbf{57}, 2968 (1998).

\bibitem {Matsumura2017}T. Matsumura, Y. Kita, Y. Yoshikawa, S. Michimura, T.
Inami, Y.Kousaka, K. Inoue, and S. Ohara, J. Phys. Soc. Jpn. \textbf{86},
124702 (2017).

\bibitem {Shibata2015}K. Shibata, J. Iwasaki, N. Kanazawa, S. Aizawa, T.
Tanigaki, M. Shirai, T. Nakajima, M. Kubota, M. Kawasaki, H. S. Park, D.
Shindo, N. Nagaosa and Y. Tokura, Nat. Nanotechnol. \textbf{10}, 589 (2015).

\bibitem {Mito2015}M. Mito, T. Tajiri, K. Tsuruta, H. Deguchi, J. Kishine, K.
Inoue, Y. Kousaka, Y. Nakao, and J. Akimitsu, J. Appl. Phys. \textbf{117},
183904 (2015).

\bibitem {Chernyshov2015}V. Dyadkin, F. Mushenok, A. Bosak, D. Menzel, S.
Grigoriev, P. Pattison, and D. Chernyshov, Phys. Rev. B \textbf{91}, 184205 (2015).

\bibitem {Togawa2015}Y. Togawa, T. Koyama, Y. Nishimori, Y. Matsumoto, S.
McVitie, D. McGrouther, R. L. Stamps, Y. Kousaka, J. Akimitsu, S. Nishihara,
K. Inoue, I.G. Bostrem, Vl. E. Sinitsyn, A.S. Ovchinnikov, and J. Kishine,
Phys. Rev. B \textbf{92}, 220412(R) (2015).

\bibitem {Wang2017}L. Wang, N. Chepiga, D.-K. Ki, L. Li, F. Li, W. Zhu, Y.
Kato, O.S. Ovchinnikova, F. Mila, I. Martin, D. Mandrus, and A.F. Morpurgo,
Phys. Rev. Lett. \textbf{118}, 257203 (2017).
\end{thebibliography}
\end{document}